How Does the Vulnerability of an Evolving Power Grid Change?

Bálint Hartmann*

Centre for Energy Research, Eötvös Loránd Research Network, KFKI Campus, Konkoly-Thege Miklós út 29–33., 1121 Budapest, Hungary

* Corresponding author. Tel.: +36 20 4825310. *E-mail address*: hartmann.balint@ek-cer.hu

**Abstract**

Topological robustness and fault tolerance of the power system are among the most studied topics when it comes to the assessment of grid vulnerability. The use of adequate methods and performance metric can greatly contribute to a number of actions, including operation planning, grid development and other specific measures. A shortcoming of several studies is that they fail to take the aspects of network evolution into consideration. In this paper the seven-decade historical dataset (1949–2019) of the Hungarian power system is used to perform vulnerability assessment applying a complex network approach. Damage tolerance of the network is examined against node and edge removals. The results show that the evolving grid has increased its tolerance against large disturbances very early, but vulnerability values show little variance from 1979, despite that the size of the network has increased significantly. It is also found that more efficient and more robust topologies impose slightly conflicting conditions for grid development planning.

# 1. Introduction

Billions of electrons have passed through the electricity networks of Hungary since 1949 when the synchronous operation of the Hungarian power system began. In that year, the first entry was written in the operations log of the National Load Dispatch Centre, the predecessor of MAVIR Hungarian Independent Transmission Operator Company Ltd., the national transmission system operator (TSO). More than seven decades have elapsed since, during which a remarkable evolution of processes, equipment and technology was seen. What was consistent though is the reliable, efficient and secure operation of the Hungarian power system, which was ensured by the TSO. This dedicated work included the continuous development of the high-voltage network, to create and maintain a grid that is less vulnerable to disturbances and deliberate acts. Illustrating this robustness by numbers, the annual amount of electricity not served due to outages of the high-voltage network was between 10 and 25 MWh during the last decade, which is negligible compared to the total consumption of approx. 34–39 TWh [1].

Being prepared to unexpected events has also gained importance in energy policy lately. The strive for legal harmonization on the European level resulted in a number of regulations, of which [2] has defined the methodology for assessing the relevance of assets for outage coordination. This methodology among other things should be based on the systematic relevance of all grid elements located in a transmission system, which necessitates adequate modelling and simulation of grid vulnerability. Another European regulation [3] was issued in 2019 on risk preparedness. This regulation has taken an extra step in order to coordinate resources better to avoid disturbances of the power system and to prepare for crisis situations. Member States of the European Union were also required to create national risk preparedness plans. The work presented in this paper aims to contribute to these activities from a methodological perspective.

Vulnerability of the power grid has gained significant attention in the literature as well in the last decade, by studies focusing on modelling aspects, fault tolerance, performance metrics and countermeasures, to name a few topics. the present paper does not intend to provide a comprehensive overview of those studies, but recommends two recent review papers, published by Pagani and Aiello [4] and Cuadra et al. [5]. The reference list of the latter includes over 200 papers, providing also an outstanding starting point for professionals who would take a deep dive into these topics. Cuadra et al. group the studies on vulnerability based on the approaches taken by the researchers. The first group includes the ones that only consider the structural aspects and use a complex network (CN) approach, while papers in the second group enhance this approach by taking into consideration models and metrics from power engineering. This paper belongs to the first group, relying dominantly on CN tools and metrics. (It also has to be noted that there is no clear agreement on whether the use of CN approaches is able to provide a full understanding of vulnerability and robustness issues of the power system [6].)

Considering the damage tolerance of the grid, papers either examine the removal of nodes [7]–[17], or the removal of edges [18]–[26]. Significantly fewer papers considered both methods [27]–[32], thus the present work contributes to the less studied parts of this topic. Of the various performance metrics (e.g. efficiency, source-demand considered efficiency, largest component size, connectivity level, clustering coefficient, power supply), efficiency was chosen as the characterizing metric of grid vulnerability. The main reason for this decision was that very scarce data is available from the first decades of the synchronous operation, which would make a proper comparison of different grid states nearly impossible, hijacking the main focus of the work (examining the aspect of grid evolution). Network efficiency has been used lately by authors to examine the Italian power grid [8],[33], the North American power grid [9], the European power grid [20], IEEE networks [21],[23], a real high-voltage power grid in Italy [22], to name a few. In terms of performance metrics, the study presented here is the most closely related to these papers.

The structure of the paper is as follows. Section 2 presents the database of the Hungarian power system and the methods that were used. In Section 3, results are presented and discussed from various aspects. Finally, conclusions are drawn in Section 4.

**2. Methods and data**

*2.1. Network data*

A network database of the Hungarian power system was created using different sources, including hand-written notes, anniversary books of utility companies, statistical publications, maps and personal consultation. The sources were not consistent, thus certain pre-processing had to be done to ensure uniform quality. In the database, a new node was created when a substation was first constructed, and a new edge was created when a power line was put into operation. If an element was decommissioned, it was removed. The final database covers 70 years and includes 400 nodes (*N*) and 774 edges (*E*).

*2.2. Methods*

To assess the vulnerability of the Hungarian power grid through its 70-year-long evolution, the CN approach was used, and converted grid topologies to graphs. This is by converting substations to nodes and power lines to edges. For the survey presented in this paper, unweighted graph representation was used, as very scarce data is available from early years, especially on line impedances or power flows, which are typically used to define weights of the edges in the graph representation. Such representations were created for each year. For the graphs, the node degree distribution and average node degree, the diameter, the modularity metric, the average path length, the clustering coefficient and the small-world metric were calculated. All calculations were carried out using MATLAB R2019a.

In the following, calculation of the CN metrics is presented.

The modularity quotient, *Q* is used to measure the strength of division of a network into modules and is defined as:

$$Q = \frac{1}{\langle k \rangle} \sum_{ij} \left( A_{ij} - \frac{k_i k_j}{\langle k \rangle} \right) \delta(g_i, g_j) \tag{1}$$

where $\langle k \rangle$ is the average node degree, $A_{ij}$ is the adjacency matrix and $\delta(i,j)$ is the Kronecker-delta function.

The average path length, *L* is the average number of steps along the shortest paths for all possible pairs of network vertices, defined as:

$$L = \frac{1}{N(N-1)} \sum_{j \neq i} d(i,j) \qquad (2)$$

where N is the number of nodes and $d(i,j)$ is the graph distance between nodes i and j.

As defined in [34], the clustering coefficient *C* is defined as:

$$C = \frac{1}{N} \sum_i \frac{2E_i}{k_i(k_i - 1)} \qquad (3)$$

where *E* is the number of edges between the neighbours of i.

When assessing the vulnerability of a network, it is also worth checking whether it shows scale-free properties. The main reason for this is that scale-free networks are more robust to random node removals, but they are more vulnerable to targeted actions (e.g. attacks) [35]–[37]. Power grids potentially showing scale-free properties were first reported by Barabási and Albert [38]. But it has been contradicted by many papers since, first by Amaral et al., who noted [39] that the degree distribution of power grids is better fitted by an exponential distribution than with a power-law, especially in the case of lower degree nodes. Aging and the limited capacity of nodes were named as potential causes of this difference. This finding was confirmed by [7]. To check whether the power system under examination displays scale-free behaviour, fitted both exponential and power-law distributions were fitted to the node degree distribution, using least squares.

As discussed in Section 1, efficiency was used as the characterizing performance metric. It is a measure of the network's performance, assuming that the efficiency for transmitting electricity between nodes *i* and *j* is proportional to the reciprocal of their distance.

$$eff = \frac{1}{N(N-1)} \sum_{j \neq i} \frac{1}{d(i,j)} \qquad (4)$$

Vulnerability of the network is defined as the drop in efficiency (*Δeff*) when a node or an edge is removed from the network. There are also different options for removal, as it can be done simultaneously as in [40],[41], or sequentially as in [41]. In this paper, simultaneous removal was chosen, mainly due to the cases with a high number of removed elements. For each examined year (1949, 1959, …, 2019) 12 scenarios were defined, in which 1, 2, 5, 10, 15 and 20 nodes or edges were removed from the network, respectively. The removed elements were selected randomly, and each scenario was run 10 000 times, which provided good statistical representation. The efficiency of the network was calculated twice, once in the initial state of the network, and once after the removal of elements. Vulnerability was determined as the ratio of efficiency decrease and initial efficiency ($\Delta eff/eff_0$). All calculations were carried out using MATLAB R2019a.

**3. Results and discussion**

*3.1. Network properties*

The results of the long-term network analysis are shown in Fig. 1. In general, it can be observed that most properties only exhibit variations in the first two decades of the evaluated period, while from 1970 on, the majority of the values can be handled as constants, as discussed in the following. The average node degree of the network ($\langle k \rangle$) varies between 1.8 and 2.61, the values marking the first and the last year of the evaluated period. The value first exceeds 2.5 in 1969 and remains practically constant for the next 50 years. The diameter of the network ($d$) is the lowest in 1949 (5) and the highest in 1970 (18). Its value varies between 14 and 15 since 1978. The modularity quotient ($Q$) of the network is in the same range since 1971 (between 0.453 and 0.469). The average path length ($L$) shows a very fast increase in the first two decades, which later transforms to a less steep but still steady trend. A similar pattern is seen when evaluating the distribution of $L$, shown in Fig. 2. In the first decade of the evolution of the network, the clustering coefficient ($C$) differs from zero only once. In the next two decades, four important increases can be seen, between 1958–1961, 1965–1969, 1977–1978 and 1985–1990. These increases can be connected to well-identifiable network development activities.

Exponential fit to the cumulative node degree distribution for 10-year snapshots are shown in Fig. 3. Parameters of the fitted distributions show little variance from 1979, despite that the size of the network has increased from 220 nodes and 281 edges to 385 nodes and 504 edges. The numerical values of the fits were compared to the ones reported in the literature and are in the same range. It can be concluded that the node distribution of the examined long-term model does not show scale-free behaviour, thus it is expected to be more vulnerable to random node/edge removals.

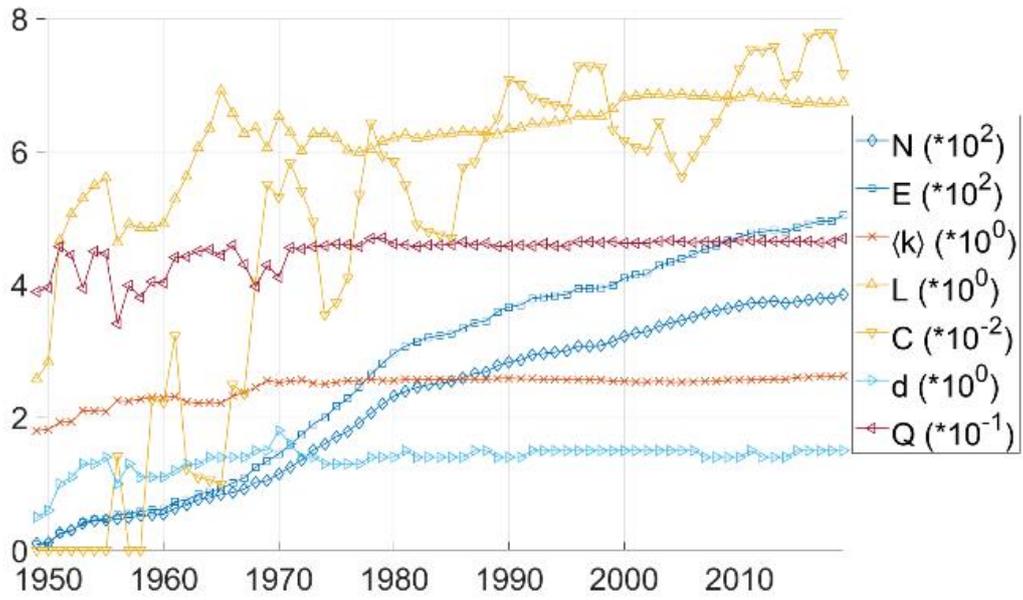

**Fig. 1. Network properties on the long-term**

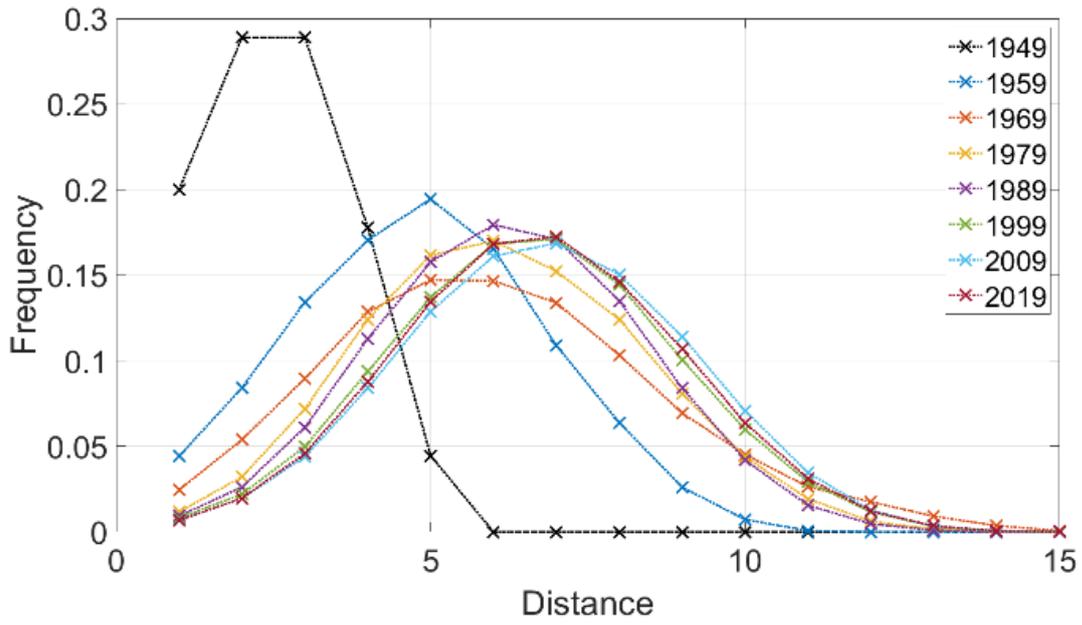

**Fig. 2. Distribution of average path length**

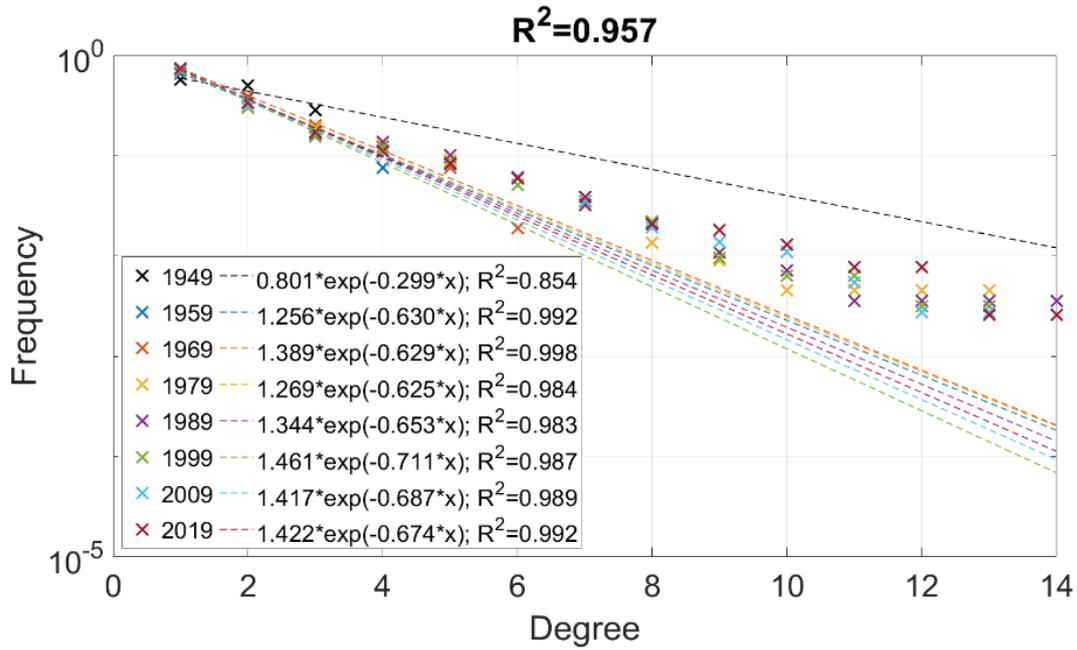

**Fig. 3. Cumulative node degree distribution and exponential fits**

It was also examined; the removal of which node/edge causes maximal damage to the grid. In the case of nodes, the most damaging element was a different one in 1949, 1959 and 1969, but remained the same since 1979. (This node represents the substation of Sajószöged, which is the largest open-air substation of Central-Eastern Europe and connects 400, 220 and 120 kV networks.) In contrast, the most damaging edge was a different one in each year. Detailed results of the vulnerability test are presented in the following sections, first according to the year of the examination, then according to the severity of the removal.

*3.2. Vulnerability in different periods of network evolution*

Table 1 shows how the ratio of disconnected edges varies as the result of removing the maximal (20) and minimal (1) number of nodes that were examined. It is seen that for the removal of 20 nodes, the level of vulnerability of the network improves very fast and reaches a decent level after four decades. A similar trend is shown after removing 1 node, it takes only three decades of evolution to reach ratios below 1%. The $k_{20}/k_1$ ratio suggests that the overall vulnerability of the grid has decreased regardless of the size of disturbance. These results are also presented in Fig. 4.

**Table 1. Ratio of edges disconnected given *k* nodes removed**

| Year | $k_{20}$ | $k_1$ | $k_{20}/k_1$ | Year | $k_{20}$ | $k_1$ | $k_{20}/k_1$ |
|---|---|---|---|---|---|---|---|
| 1949 | 0.7500* | 0.0833 | 9 | 1989 | 0.1678 | 0.0071 | 23.5 |
| 1959 | 0.6 | 0.0181 | 33 | 1999 | 0.1518 | 0.0063 | 24 |
| 1969 | 0.3925 | 0.0186 | 21 | 2009 | 0.1311 | 0.0054 | 24 |
| 1979 | 0.2117 | 0.009 | 23.5 | 2019 | 0.1188 | 0.0051 | 23 |
| *15 nodes | | | | | | | |

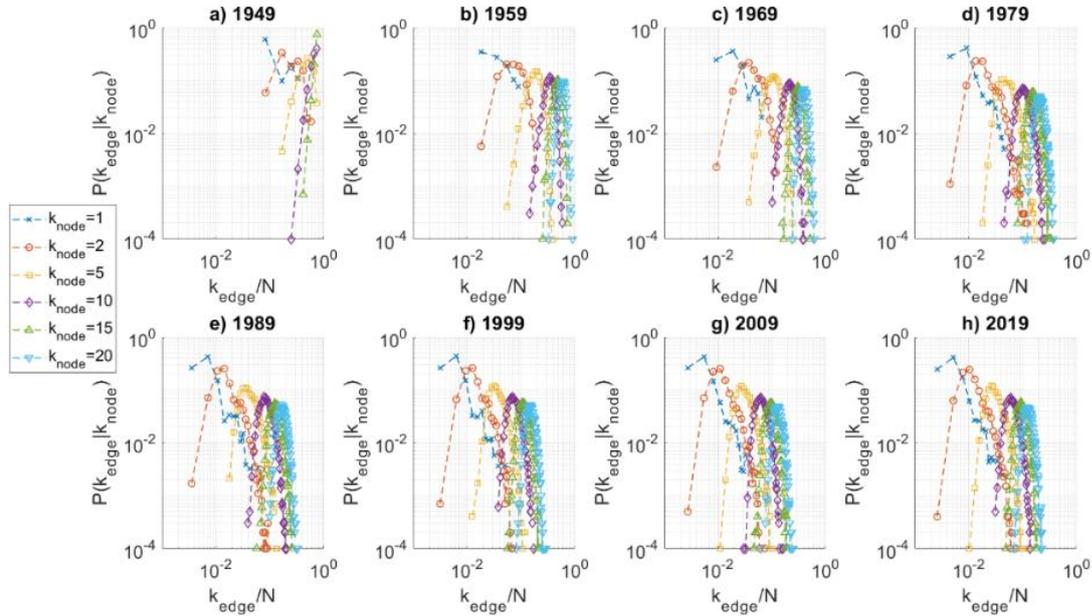

**Fig 4. Ratio of edges disconnected given *k* nodes removed in different years**

Table 2 shows the maximal and most probable values of grid vulnerability, given *k* nodes removed from the network. It is seen that the removal of a single node causes negligible disturbances from as early as 1969, while the maximal vulnerability for single node removal remains unchanged after 1979. It suggests that although the size of the network has increased by 3.5 times between that period and today, damage tolerance did not change. The $k_{20}/k_1$ ratio shows a decreasing trend, which implies that tolerance against larger disturbances has improved. If one compares the ratio of the maximal and the most probable values for single node removal, an increasing trend is seen between 1949 and 1979 (1→16), but a decreasing one after 1979 (16→8), thus the most vulnerable node of the grid was installed by 1979. This is the substation of Sajószöged, as mentioned earlier, the importance of which could only be partially reduced by four decades of network development. These results are presented in Fig. 5.

**Table 2. Maximal and most probable values of $\Delta eff_-/eff_0$ given $k$ nodes removed**

| Year | Maximal values | | | Most probable values | | |
|---|---|---|---|---|---|---|
| | $k_{20}$ | $k_1$ | $k_{20}/k_1$ | $k_{20}$ | $k_1$ | $k_{20}/k_1$ |
| 1949 | 0.9900* | 0.3960 | 2.5 | 0.8316* | 0.1683 | 4.94 |
| 1959 | 0.7722 | 0.0792 | 9.75 | 0.5742 | 0.0198 | 29 |
| 1969 | 0.5049 | 0.0297 | 17 | 0.1980 | 0.0099 | 20 |
| 1979 | 0.2673 | 0.0198 | 13.5 | 0.0891 | 0.0099 | 9 |
| 1989 | 0.1881 | 0.0198 | 9.5 | 0.0792 | 0.0099 | 8 |
| 1999 | 0.1683 | 0.0198 | 8.5 | 0.0693 | 0.0099 | 7 |
| 2009 | 0.1386 | 0.0198 | 7 | 0.0594 | 0.0099 | 6 |
| 2019 | 0.1188 | 0.0198 | 6 | 0.0495 | 0.0099 | 5 |
| *15 nodes | | | | | | |

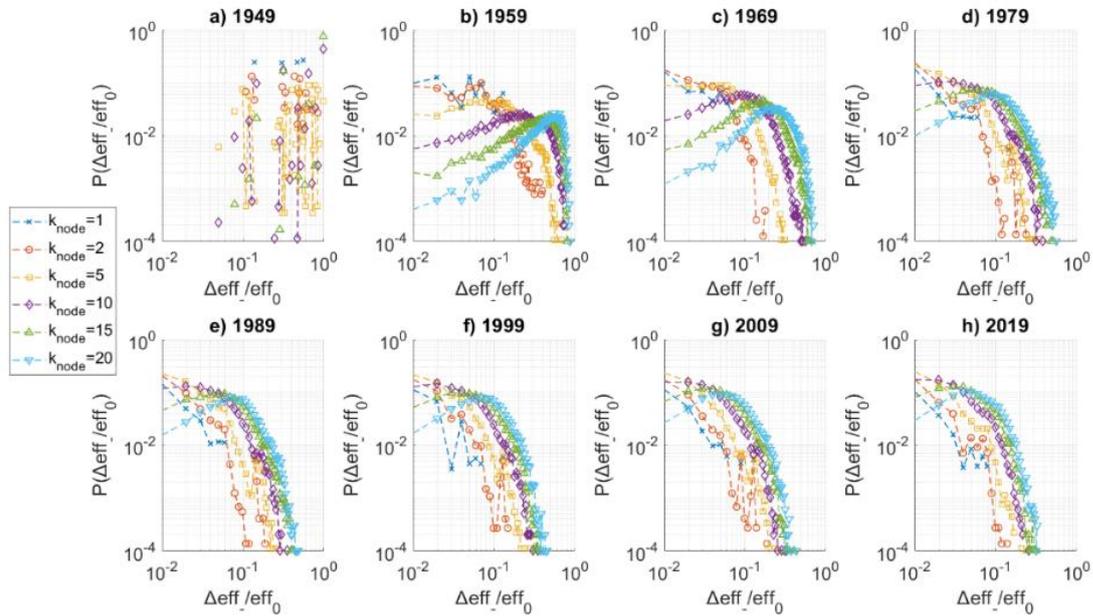

**Fig 5. Decrease of network efficiency given $k$ nodes removed in different years**

Table 3 shows the maximal and the most probable values of grid vulnerability, given k edges removed from the network. It is noticeable that values of the maximal and the most probable damage is smaller by approx. an order of magnitude compared to values in Table II, which means that the network is far less vulnerable to edge removal than to node removals. Damage caused by single edge removals reaches the minimum from 1969–1979, and the $k_{20}/k_1$ ratio decreases continuously. The ratio of maximal and most probable values for single edge removal are constant since 1979, which shows that there are no significant weak connections in the system. These results are also presented in Fig. 6.

**Table 3. Maximal and most probable values of $\Delta eff./eff_0$ given $k$ edges removed**

| Year | Maximal values | | | Most probable values | | |
|---|---|---|---|---|---|---|
| | $k_{20}$ | $k_1$ | $k_{20}/k_1$ | $k_{20}$ | $k_1$ | $k_{20}/k_1$ |
| 1949 | 0.9900* | 0.3960 | 2.5 | 0.8316* | 0.1683 | 4.94 |
| 1959 | 0.7722 | 0.0792 | 9.75 | 0.5742 | 0.0198 | 29 |
| 1969 | 0.5049 | 0.0297 | 17 | 0.1980 | 0.0099 | 20 |
| 1979 | 0.2673 | 0.0198 | 13.5 | 0.0891 | 0.0099 | 9 |
| 1989 | 0.1881 | 0.0198 | 9.5 | 0.0792 | 0.0099 | 8 |
| 1999 | 0.1683 | 0.0198 | 8.5 | 0.0693 | 0.0099 | 7 |
| 2009 | 0.1386 | 0.0198 | 7 | 0.0594 | 0.0099 | 6 |
| 2019 | 0.1188 | 0.0198 | 6 | 0.0495 | 0.0099 | 5 |
| *15 edges | | | | | | |

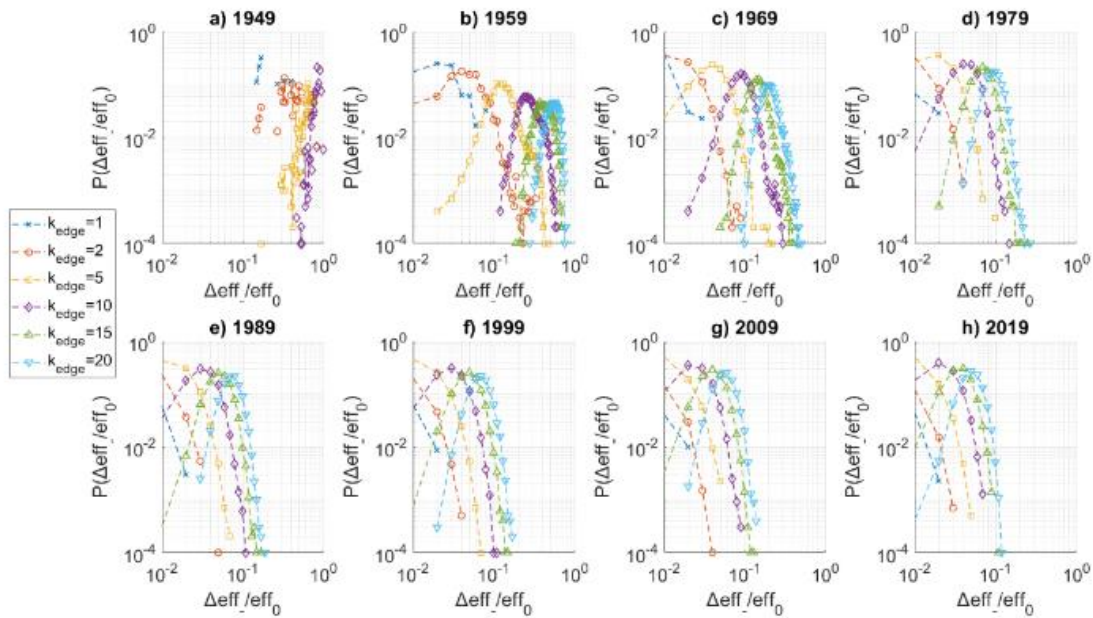

**Fig 6. Decrease of network efficiency given $k$ edges removed in different years**

*3.3. Vulnerability in case of different removals*

Table 4 shows how the ratio of disconnected edges varies between 1949 and 2019 as the result of removing a different number of nodes. The most important finding is that while seven decades of grid development have significantly improved robustness against small disturbances, the simultaneous outage of 5+ nodes would still cause a drop in efficiency over 10%. These results are presented in Fig. 7.

**Table 4. Ratio of edges disconnected given $k$ nodes removed**

| Node | 1 | 2 | 5 | 10 | 15 | 20 |
|---|---|---|---|---|---|---|
| 1949 | 0.3333 | 0.5833 | 0.7500 | 0.7500 | 0.7500 | 0.9090* |
| 2019 | 0.0413 | 0.0749 | 0.1266 | 0.1421 | 0.1937 | 0.2299 |
| 1949/2019 | 8.0702 | 7.7877 | 5.9241 | 5.2779 | 3.8719 | 3.9538 |
| *1959 | | | | | | |

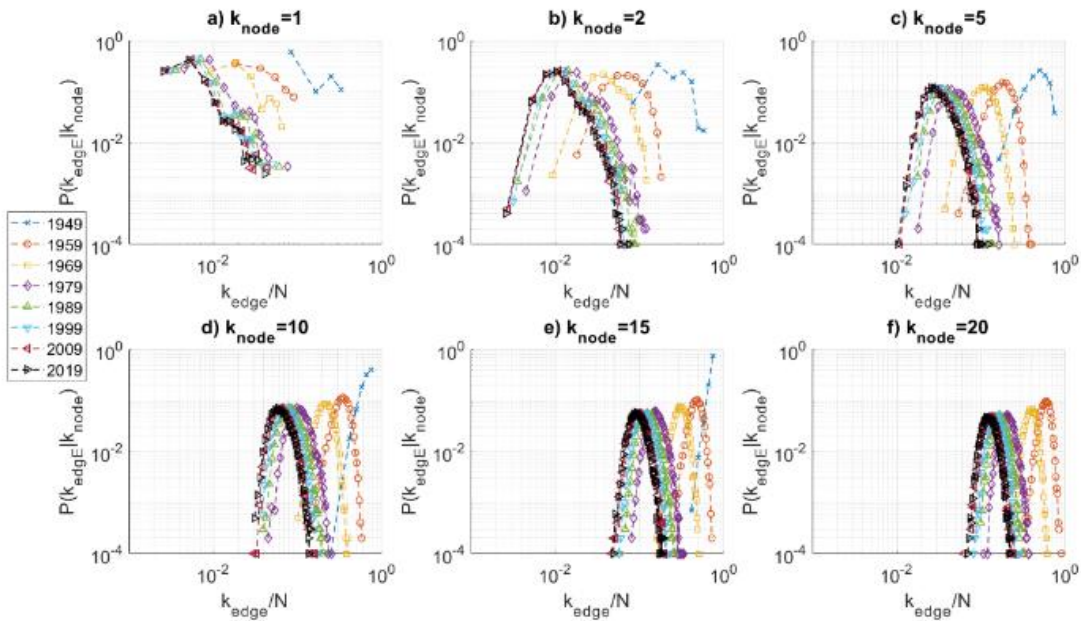

**Fig 7. Ratio of edges disconnected given *k* nodes removed in the case of different removals**

Table 5 shows the maximal and most probable values of grid vulnerability, given *k* nodes removed from the network. Considering maximal values, it is seen that even today one is able to select 2 nodes which results in a ~14% drop in efficiency if removed. Also, the vulnerability against removing more than 5 nodes did not decrease that significantly. In contrast, the network has become very robust against random node removals. These altogether imply good connectivity, and the existence of a handful of nodes with large degree (as in scale-free networks). These results are also presented in Fig. 8.

**Table 5. Maximal and most probable values of *Δeff./eff$_0$* given *k* nodes removed**

| Node | 1 | 2 | 5 | 10 | 15 | 20 |
|---|---|---|---|---|---|---|
| Maximal values | | | | | | |
| 1949 | 0.5544 | 0.8415 | 0.9900 | 0.9900 | 0.9900 | 0.8712* |
| 2019 | 0.0792 | 0.1386 | 0.2673 | 0.2574 | 0.3168 | 0.3267 |
| 1949/2019 | 7 | 6.0714 | 3.7037 | 3.8461 | 3.125 | 2.666 |
| Most probable values | | | | | | |
| 1949 | 0.5544 | 0.4356 | 0.6534 | 0.9900 | 0.9900 | 0.5643* |
| 2019 | 0.0099 | 0.0099 | 0.0099 | 0.0099 | 0.0297 | 0.0396 |

| 1949/2019 | 56 | 44 | 66 | 100 | 33.3333 | 14.25 |
| *1959 | | | | | | |

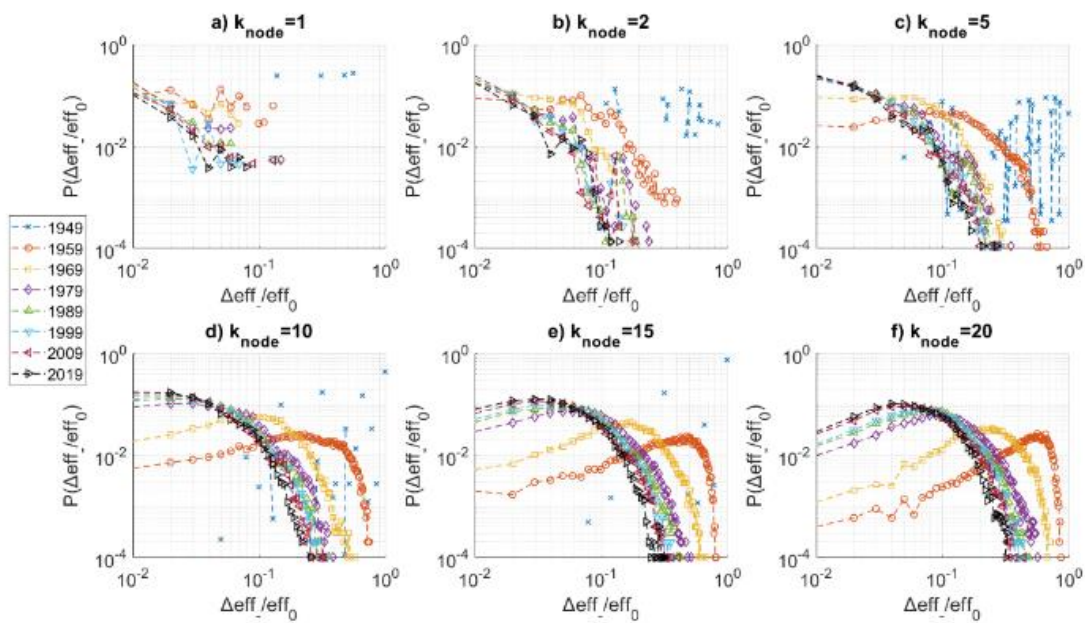

**Fig 8. Decrease of network efficiency given *k* nodes removed in the case of different removals**

Table 6 shows the maximal and most probable values of grid vulnerability, given *k* edges removed from the network. The results show a very similar pattern to node removals, but once again higher robustness is seen. To reach a 10% drop in efficiency, 15 of 504 lines have to be removed in a coordinated manner. These results are presented in Fig. 9.

**Table 6. Maximal and most probable values of *Δeff./eff₀* given *k* edges removed**

| Edge | 1 | 2 | 5 | 10 | 15 | 20 |
|---|---|---|---|---|---|---|
| Maximal values | | | | | | |
| 1949 | 0.3960 | 0.5346 | 0.8118 | 0.9900 | 0.6831* | 0.7722* |
| 2019 | 0.0198 | 0.0297 | 0.0495 | 0.0693 | 0.1089 | 0.1188 |
| 1949/2019 | 20 | 18 | 16.4 | 14.2857 | 6.2727 | 6.5 |
| Most probable values | | | | | | |
| 1949 | 0.1683 | 0.3267 | 0.6435 | 0.8316 | 0.3663* | 0.5742* |
| 2019 | 0.0099 | 0.0099 | 0.0099 | 0.0198 | 0.0396 | 0.0495 |
| 1949/2019 | 17 | 33 | 65 | 41 | 9.25 | 11.6 |
| *1959 | | | | | | |

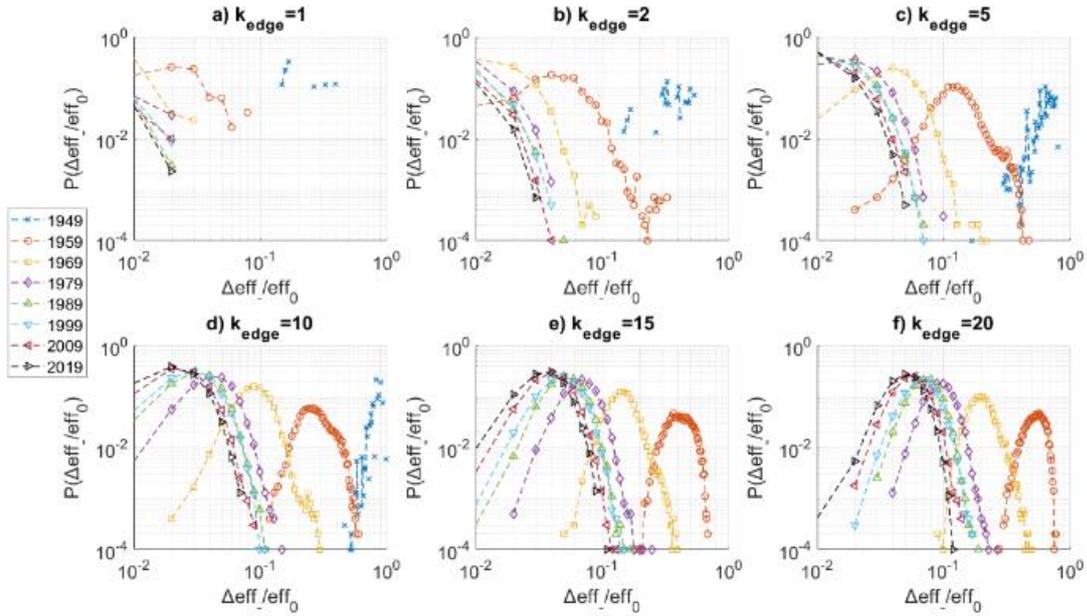

**Fig 9. Decrease of network efficiency given *k* edges removed in the case of different removals**

*3.4. Discussion*

Based on the results, a main finding of this survey is that the vulnerability of the Hungarian power system against the removal of a single node/edge did not improve significantly since 1979. (This date practically also marks the end of the large-scale development of the 220 and 400 kV network, which took place in two periods. Between 1965 and 1969 7 220 kV lines were commissioned: Dunamenti–Zugló, Dunamenti–Soroksár, Sajószöged–Szolnok, Zugló–Göd, Sajószöged–Detk, Detk–Zugló and Detk–Szolnok. The backbone of the 400 kV network was constructed between 1978 and 1979, with 9 lines: Tisza Power Plant–Sajószöged, Albertfalva–Dunamenti, Sajószöged–Göd, Sajószöged–Felsőzsolca, Dunamenti–Martonvásár, Litér–Martonvásár, Martonvásár–Toponár and Győr–Litér. It was also displayed that while the node degree distribution of the network did not suggest scale-free behaviour, the grid does have a number of nodes of extreme importance. The removal of these nodes can lead to damage larger than typical, which is characterized by many studies as a sign of scale-free behaviour. Reconsideration of these aspects might be necessary. The results also confirm the findings of [42], who stated that efficient networks have a multi-hubs star-like structure, which is fragile for the removal of high-degree nodes. This, however, raised another question for future work: should power grid development aim for more efficient or for more robust network topologies, as their requirements seem to be slightly conflicting.

It was also examined how certain CN parameters could be linked to the evolution of vulnerability. To do so, maximal values of $\Delta eff_-/eff_0$ were normalised for each year, both for node and edge removals, and the average of seven decades was plotted (Fig. 10 and 11, respectively). The average path length and the clustering coefficient of the network were also normalised and plotted. It can be seen that the CN parameters show a very similar trend to the maximal values of damage. The correlation of average damage and $L$ was -0.9880 and -0.9820 for node and edge removal, respectively. The correlation of average damage and $C$ was -0.9892 and -0.9924 for node and edge removal, respectively. Lower, but still strong correlation was measured with the small-world coefficient (-0.86). These results imply that the maximal decrease of efficiency for a given network and given damage is largely related to certain CN parameters, which could contribute to the findings of [6], mentioned in Section 1.

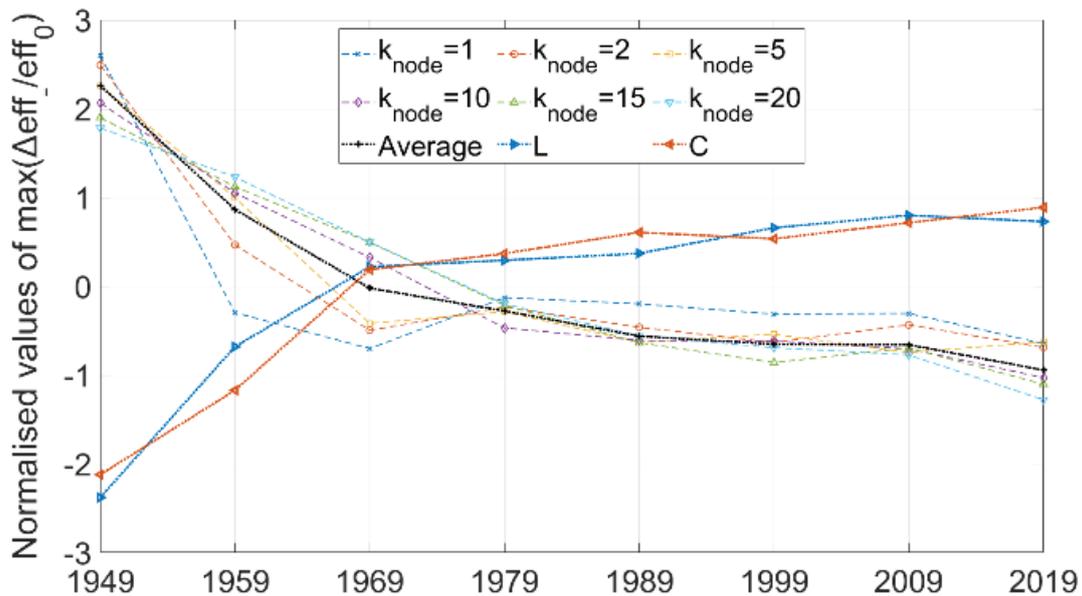

**Fig 10. Normalised values of $\Delta eff_-/eff_0$ and selected CN parameters, node removal**

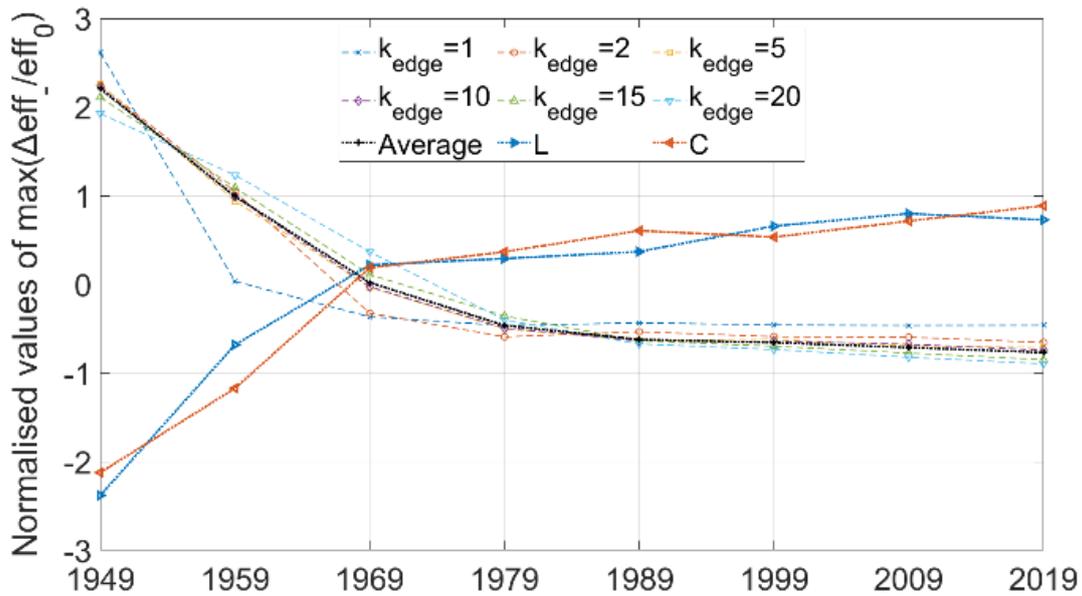

**Fig 11.** Normalised values of *Δeff./eff₀* and selected CN parameters, edge removal

## 4. Conclusions

In this paper the vulnerability of the Hungarian power system was studied using a complex network approach. Topological efficiency of the grid was calculated against node and edge removals of different size, between 1 and 20 elements. Maximal and most probable values of efficiency decrease were evaluated for different periods of network evolution and for different sizes of removals. The results show that the evolving grid has increased its tolerance against large disturbances very early. Vulnerability values show little variance from 1979, despite that the size of the network has increased significantly. This suggests on the one hand that the elements most important for security have been installed before this date, but on the other hand it implies that grid development of the last 40 years could not significantly contribute to this aspect. It was shown that more efficient and more robust topologies impose slightly conflicting conditions for grid development planning. It was also shown that the evolution of the average path length and the clustering coefficient show very strong correlations with maximal damage, regardless of the size of the node/edge removal.

**Acknowledgement**

The author would like to thank the following people for their contribution to assembling the historical grid database: Viktória Sugár, Attila Dervarics, Zoltán Feleki, József Hiezl, János Nemes, Imre Orlay, János Rejtő.

**Declaration of Competing Interest**

The author declares that he has no known competing financial interests or personal relationships that could have appeared to influence the work reported in this paper.